\documentclass{article}
\PassOptionsToPackage{numbers}{natbib}
\usepackage[preprint]{neurips_2026}

\usepackage[utf8]{inputenc} 
\usepackage[T1]{fontenc}    
\usepackage{hyperref}       
\usepackage{url}            
\usepackage{booktabs}       
\usepackage{amsfonts}       
\usepackage{nicefrac}       
\usepackage{microtype}      
\usepackage{xcolor}         
\usepackage{microtype}
\usepackage{graphicx}
\usepackage{subcaption}
\usepackage{times}
\usepackage{latexsym}
\usepackage{tabularx}
\usepackage{amsfonts}
\usepackage{multirow}
\usepackage{enumitem}
\usepackage{makecell}
\usepackage{soul}
\usepackage{url}
\usepackage{algorithm}
\usepackage{algorithmic}
\usepackage{bm}
\usepackage{amsmath}
\usepackage{amssymb}
\usepackage{mathtools}
\usepackage{amsthm}
\usepackage{hyperref}

\title{Adaptive Compression and Targeted Perturbation:\\
A Unified Framework for\\
Generalized Audio Deepfake Detection}

\author{%
  Jianqiao Cui\thanks{Corresponding authors.} \\
  Yangtze Delta Region Institute of Tsinghua University\\
  Jiaxing, China\\
  \texttt{jc18n17@soton.ac.uk} \\
  \And
  Haoxian Zhu \\
  University of Southampton \\
  Southampton, UK \\
  \texttt{h2v20@soton.ac.uk} \\
  \AND
  Hanbiao Meng \\
  Taiyuan University of Technology\\
  Taiyuan, China\\
  \texttt{2024001543@link.tyut.edu.cn} \\
  \And
  Yanghao Zhang\footnotemark[2]  \\
  Imperial College London \\
  London, UK \\
  \texttt{yanghao.zhang@imperial.ac.uk} \\
}

\begin{document}

\setcounter{footnote}{1}
\maketitle
\setcounter{footnote}{0}

\begin{abstract}
Existing audio deepfake detection (ADD) methods frequently exhibit limited generalization, as fixed bottleneck architectures fail to adapt to the heterogeneous information densities inherent in diverse spoofing artifacts. Moreover, standard augmentation often leads to a pathological robustness-accuracy trade-off. In this paper, we propose a unified ADD framework to address these challenges. Our approach integrates: (1) an Adaptive Audio Feature Integration (AAFI) module that dynamically selects optimal latent dimensions via a compression pool and a stochastic exploration mechanism, ensuring flexible representations across multiple types of dynamic attacks; and (2) a Mel-Band Adversarial Perturbation (Mel-BAP) strategy that applies targeted regularization in the mel-spectrogram domain to encourage the model to learn invariant decision boundaries. 
Extensive evaluations on benchmarks including ADD 2022, FakeOrReal, and In-the-Wild demonstrate that our model based on Whisper-small achieves a state-of-the-art average Equal Error Rate (EER) of 1.51\%. Compared to existing SOTA models such as the 3-billion-parameter Resemble-Detect-3B-Omni, our framework reduces the parameter count by up to 91.6\% while achieving a 60.7\% improvement in average EER. Notably, the model exhibits exceptional zero-shot transferability, achieving EERs of 3.91--5.65\% on unseen languages and maintaining strong resilience to real-world acoustic distortions. This work provides a scalable and efficient “plug-in” solution for enhancing trust and security in the audio Web ecosystem.
\end{abstract}

\section{Introduction}

Rapid advancements in speech generation \cite{li2025survey,triantafyllopoulos2023overview,ye2024flashspeech} have enabled the integration of natural-sounding voices into virtual assistants and content creation \cite{tan2021survey,ren2020fastspeech}. However, the misuse of these technologies through audio \textit{deepfakes} poses severe security and privacy threats to the Web ecosystem, facilitating the spread of misinformation and causing reputational damage \cite{chintha2020recurrent,zhang2020learning,wu2023defender}. Consequently, robust detection methods are essential to maintain digital trust.
While initiatives such as ASVspoof \cite{todisco2019asvspoof,nautsch2021asvspoof} and ADD \cite{yi2022add} have catalyzed algorithm development, cross-domain generalization remains a critical challenge. Detection models often suffer substantial performance degradation on unseen deepfakes \cite{yamagishi2021asvspoof} and across heterogeneous evaluation conditions \cite{muller2022does}. Although models built on pre-trained speech representations, such as Wav2vec \cite{huang2025speechfake}, have shown promising performance \cite{huang2025speechfake,li2024safeear,zhang2024deepfake,cui2025whiadd}, most apply the same fixed-dimensional transformation to every input, despite substantial variation in the salience and spectro-temporal complexity of spoofing artifacts. From the perspective of the trade-off between compression and fidelity associated with information bottleneck theory \cite{chalapathy2019learning,kawaguchi2023how}, narrower representations may suppress task-irrelevant variation but risk discarding weak forensic cues, whereas wider representations may preserve subtle artifacts while retaining redundant or noisy information. A fixed representation width may therefore be suboptimal across the diverse noise levels, synthesis methods, and attack distributions encountered in practice \cite{wang2023drift,liu2024unitime,martindonas2022vicomtech,guo2024deepfake}. This motivates input-dependent multi-resolution representations that complement the original pre-trained features and adapt their compression level to heterogeneous spoofing conditions. Beyond unseen spoofing methods, adversarial perturbations pose an additional challenge to neural network robustness \cite{zhang2023guap}. Prior work has explored adversarial training strategies that address training efficiency \cite{wang2023deat}, adaptivity \cite{wang2023seat}, and class-wise robustness disparities \cite{zhang2024faal}.

To address these challenges, we investigate the following research questions:


\vspace{1mm} 
\noindent\fbox{\parbox{0.97\linewidth}{
\small \textit{RQ1:} Can input-dependent compressed representations complement frozen encoder features across heterogeneous spoofing conditions? \quad 

\textit{RQ2:} How can we enhance robustness against unseen spoofing methods and adversarial attacks?
}}
\vspace{1mm}

To investigate \textit{RQ1}, we introduce the \underline{A}daptive \underline{A}udio \underline{F}eature \underline{I}ntegration (AAFI) module. Rather than applying a fixed-width transformation to every input, AAFI uses an input-dependent router to select and fuse complementary representations from a heterogeneous compression pool. These multi-resolution features are combined with the original Whisper encoder representation to accommodate diverse spoofing conditions \cite{dong2023distdet} while limiting information loss.

To enhance robustness (\textit{RQ2}), we propose the \textit{Mel-Band Adversarial Perturbation (Mel-BAP)} strategy. Recognizing that global augmentation can clutter the information bottleneck and sacrifice precision, we design Mel-BAP to perform adversarial training directly in the mel-spectrogram domain. By leveraging perceptually scaled frequency bands, it encourages the network to learn robust, perceptually aligned decision boundaries resilient to unseen spoofing patterns. We integrate AAFI and Mel-BAP into a unified framework and make the following contributions:
\begin{itemize}[leftmargin=*, itemsep=0pt, parsep=0pt]
    \item \textbf{AAFI Module:} An adaptive integration mechanism with a compression pool to handle diverse information densities in dynamic attacks.
    \item \textbf{Mel-BAP Strategy:} A targeted frequency-domain adversarial training approach. Unlike generic frequency-selective methods, Mel-BAP is perceptually aligned with the Whisper backbone and exploits forensic priors associated with neural vocoders to achieve robustness.
    \item \textbf{State-of-the-art Performance:} Extensive tests on the ADD 2022, FakeOrReal, and In-the-Wild datasets show that our method outperforms current SOTA approaches by 28.8\% with fewer trainable parameters.
\end{itemize}

\section{Related Work}
\paragraph{Audio Deepfake Datasets.} We leverage two large-scale datasets: SpeechFake \cite{huang2025speechfake}, a multilingual corpus covering 46 languages and 40 generation methods, and CodecFake \cite{wu2024codecfake2}, which focuses on neural codec-based forgeries. These datasets provide the diversity necessary to evaluate cross-lingual and cross-method generalization. Detailed statistics for both are provided in Appendix~\ref{sec:Dataset Statistics}.

\paragraph{Foundation Models for ADD.} 
Pre-trained foundation models such as Whisper \cite{radford2023robust} have shown superior robustness in audio tasks due to extensive pre-training on 680k hours of diverse data. Recent research indicates that Whisper’s self-attention architecture captures finer acoustic cues than CNN-based models do, improving cross-dataset generalization \cite{luo2024whisper}. To further validate this, we conducted an extensive cross-subset evaluation on the CodecFake dataset (see Appendix~\ref{sec:app_crosstest}), which confirms that Whisper-based models effectively capture invariant spoofing patterns across diverse neural codec artifacts, establishing a robust foundation for the proposed AAFI and Mel-BAP modules.

Unlike existing works that typically treat Whisper as a fixed feature extractor \cite{kawa2023improved}, our framework integrates the AAFI module to enable dynamic, input-tailored compression. This addresses the challenge of heterogeneous artifact densities across different forgeries, an aspect largely overlooked by static bottleneck architectures. Beyond representation learning, frequency-domain adversarial training has been explored to improve robustness in computer vision. FreqAT \cite{zhang2024freqat} combines adaptive frequency-domain transformations with adversarial training to improve model robustness while mitigating image quality degradation. Mask-constrained adversarial perturbations have also been explored in object detection \cite{zhang2021hnm}. In contrast, our Mel-BAP is designed specifically for deepfake detection. It incorporates the forensic insight that neural vocoder artifacts are disproportionately concentrated in high-frequency bands (4–8 kHz) \cite{zhang2025fsat}, allowing for more targeted and efficient regularization than existing approaches.


\section{Methodology}
\subsection{Overview}
\label{sec:overview}

Let $x$ denote an input audio waveform and $y$ its corresponding ground-truth label. In this work, we adopt the Whisper architecture, which employs a fixed pre-processing module, denoted by $\mathcal{S}(\cdot)$, to convert raw waveforms into log-Mel spectrograms $M = \mathcal{S}(x)$. For ADD tasks, neural networks are typically trained to predict the label by minimizing the cross-entropy loss $\mathcal{L}(\hat{y},y)$. Specifically, our proposed Mel-BAP strategy operates in the frequency domain by applying targeted perturbations directly to this log-Mel representation. Thus, the prediction is formulated as $\hat{y}=\mathcal{F}_{\theta}(\text{Mel-BAP}(M))$, and the model parameters $\theta$ are optimized by minimizing the following objective:
\begin{equation}
    \mathcal{L}(\mathcal{F}_{\theta}(\text{Mel-BAP}(M)), y)
\end{equation}

\paragraph{Justification for Decoder-based Classification.}
While a lightweight multilayer perceptron (MLP) head is standard for binary classification, we leverage the Whisper decoder for two strategic reasons.
First, unlike MLP heads that typically require global pooling (sacrificing temporal granularity), the decoder's \textit{cross-attention mechanism} dynamically attends to specific frames within the acoustic representation $H_{acous}$. This is critical for detecting deepfakes, where artifacts are often localized (e.g., phase discontinuities) rather than globally distributed.
Second, utilizing the pre-trained decoder preserves the semantic-acoustic alignment learned during large-scale supervision, offering better initialization than using a randomly initialized classification head.
The decoding task is simplified to generating a single class token (e.g., predicting the token for ``Genuine'' or ``Spoofing'' immediately after the start-of-transcript token), minimizing inference latency compared to full speech transcription.

\begin{figure*}[ht]
    \centering
    \includegraphics[width=1.0\textwidth]{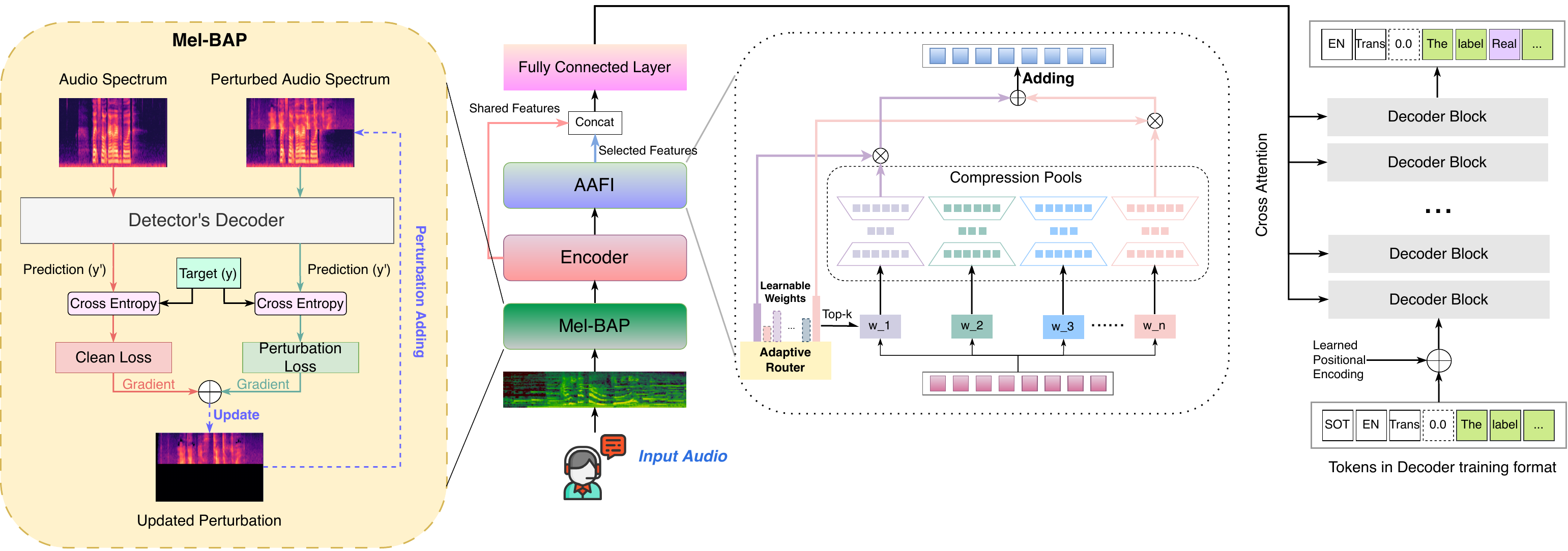}
    \caption{Schematic of the robust audio forgery detection pipeline. Left: Mel-BAP generates band-selective perturbations via PGD as targeted adversarial regularization. Center: AAFI implements an adaptive information bottleneck, using a router to dynamically select compressed features from heterogeneous pools. Right: Fusion of multi-resolution features for sequence-to-sequence decoding.}
    \label{fig:adaptive_bottleneck_arch}
\end{figure*}

Figure \ref{fig:adaptive_bottleneck_arch} illustrates the overall process, with an emphasis on the proposed Adaptive Audio Feature Integration (AAFI) module, which is positioned between the encoder and decoder components of the base model. During training, the entire model is optimized in an end-to-end manner. Crucially, to preserve the rich acoustic representations learned during large-scale pre-training and to minimize computational overhead, the parameters of the Whisper encoder remain fixed. The gradient updates are exclusively applied to the proposed AAFI module, the projection layers, and the Whisper decoder.
To improve the granularity of the downstream classifier's input, we concatenate the shared features (the output of the base model encoder) with the selected features (the output of the proposed AAFI module) and pass them through a fully connected layer to adjust their dimensionality. The adaptive routing mechanism allows gradients to flow through the selected compression pools, enabling the adaptive router to learn meaningful selection policies based on audio characteristics. The Mel-BAP component further encourages the model to learn more robust and discriminative audio features. The detailed structure of Mel-BAP is presented in the left part of Figure~\ref{fig:adaptive_bottleneck_arch}. In the following sections, we provide a detailed description of the two core components of the proposed framework.

Unlike standard classification models that rely on a simple classification head, our framework leverages the generative capabilities of the Whisper decoder. First, we integrate the diverse information captured by the framework. Let $\mathbf{Z}_{enc} \in \mathbb{R}^{T \times d_{enc}}$ denote the shared features from the fixed Whisper encoder, and $\mathbf{Z}_{aafi} \in \mathbb{R}^{T \times d_{aafi}}$ denote the selected adaptive features from the AAFI module. To preserve both general acoustic patterns and detailed artifact information, we concatenate these features along the channel dimension: $\mathbf{Z}_{cat} = \text{Concat}(\mathbf{Z}_{enc}, \mathbf{Z}_{aafi})$.
Subsequently, a fully connected (FC) layer is employed to map the concatenated features to the dimension required by the decoder ($d_{model}$): $\mathbf{H}_{acous} = \text{ReLU}(\mathbf{Z}_{cat} \mathbf{W}_{fc} + \mathbf{b}_{fc})$,
where $\mathbf{W}_{fc}$ and $\mathbf{b}_{fc}$ are learnable parameters. 
$\mathbf{H}_{acous}$ serves as the robust acoustic representation fed into the Whisper decoder. 
The decoder operates in a sequence-to-sequence manner. It receives the acoustic representation $\mathbf{H}_{acous}$ (as keys and values for cross-attention) and a sequence of prompt tokens (as queries). 
The model is trained to generate a target sequence $\mathbf{Y} = \{y_1, y_2, \dots, y_L\}$ that represents the predicted label (e.g., the text tokens for "Genuine" or "Spoofing").
The probability of the target sequence is computed autoregressively:
\begin{equation}
    P(\mathbf{Y}|\mathbf{H}_{acous}) = \prod_{t=1}^{L} P(y_t | y_{<t}, \mathbf{H}_{acous})
\end{equation}
In our ADD task, the prediction is determined by the probability assigned to the class-specific tokens in the generated sequence.

To optimize the framework end-to-end and handle data imbalance, we employ a weighted cross-entropy loss on the generated tokens. 
Let $y_{target}$ be the ground-truth token index for a given sample, and $p(y_{target})$ be the probability predicted by the decoder for that token. The loss $\mathcal{L}_{WCE}$ is formulated as:
\begin{equation}
    \mathcal{L}_{WCE} = - \frac{1}{N} \sum_{i=1}^{N} w_{y_i} \log p(y_{target}^{(i)} | \mathbf{H}_{acous}^{(i)})
\end{equation}
where $N$ is the batch size, and $w_{y_i}$ is the class weight. To ensure stable training and balanced gradient flow, we employed a hybrid approach combining oversampling with class weighting: we replicated the genuine samples (minority class) to reach an effective training ratio of 3:5 (Genuine:Spoofed). Accordingly, we set $w_{genuine}=1.0$ and $w_{spoofed}=0.6$ to prevent the model from becoming biased towards the majority class, thereby improving the detection of genuine speech (reducing the false rejection rate).
Combined with the Mel-BAP strategy, the final objective minimizes the expected loss over the perturbed inputs:


\begin{equation}
    \mathcal{L} = \mathcal{L}_{WCE}(\mathcal{F}_{\theta}(M), Y) + \mathcal{L}_{WCE}(\mathcal{F}_{\theta}(\text{Mel-BAP}(M)), Y)
\end{equation}

\paragraph{Cross-lingual Generalization Analysis.} 
Finally, to further stress-test the zero-shot transferability of our framework under extreme conditions, we evaluated its performance in five unseen languages (es, fr, ja, it, ko) from the SpeechFake MD dataset. Despite being trained only on bilingual (EN/CN) data, our model maintains robust detection capabilities across all linguistic boundaries. As detailed in Appendix \ref{sec:cross-lingual}, these results provide additional evidence that the AAFI and Mel-BAP modules effectively capture language-invariant spoofing cues, preventing the model from relying excessively on language-specific acoustic patterns.

\subsection{Adaptive Audio Feature Integration}
\label{sec:aafi}

\paragraph{Motivation.}
Audio deepfake detection (ADD) faces a fundamental challenge: the \textit{heterogeneous information density} of spoofing artifacts. High-fidelity deepfakes (e.g., generated by advanced neural codecs) exhibit subtle artifacts that require high-dimensional representations to capture, whereas low-quality spoofs often contain obvious noise, and high-dimensional features may introduce redundancy and overfitting in such cases. Existing methods typically employ fixed-size compression layers, creating a suboptimal trade-off between fidelity and compactness. 
To resolve this, we propose the \textbf{\underline{A}}daptive \textbf{\underline{A}}udio \textbf{\underline{F}}eature \textbf{\underline{I}}ntegration (\textbf{AAFI}) module. 
Unlike standard mixture-of-experts (MoE) architectures that utilize homogeneous experts to scale model capacity \cite{negroni2025leveraging}, AAFI acts as a \textit{dynamic intrinsic filter} based on the information bottleneck principle. It adaptively selects the optimal compression ratio for each input sample, maximizing the mutual information between the latent representation and the spoofing labels while minimizing irrelevant noise.


\paragraph{Heterogeneous Compression Pool.}
To accommodate the varying information densities, we construct a heterogeneous compression pool consisting of $C$ parallel branches. Unlike standard MoE experts that share identical architectures, each branch in our pool has a \textit{distinct latent dimension} $d_i \in \{d_{\min}, \dots, d_{\max}\}$. This design provides a spectrum of ``resolutions'' for the audio features. Given an encoded audio feature representation $\mathbf{Z}_{enc} \in \mathbb{R}^{T \times d_{r}}$ (where $d_{r}$ equals the encoder's hidden dimension, e.g., 768 for Whisper-small), each branch performs dimensionality reduction via $\mathbf{Z}_{i} = \text{DownNet}_{i}(\mathbf{Z}_{enc}) \in \mathbb{R}^{T \times d_{i}}, \quad i=1, \dots, C$,
where $\text{DownNet}_{i}$ is implemented as a single-layer learnable linear projection to minimize computational overhead. Detailed architectural specifications, including the exact dimensionality settings and complexity analysis, are provided in Appendix~\ref{app:aafi_details}.

\paragraph{Adaptive Routing with Exploration.}
To dynamically assign the optimal information bottleneck for each sample, we first apply temporal mean pooling to the encoded features to obtain an utterance-level representation $\bar{z} = \frac{1}{T} \sum_{t=1}^{T} \mathbf{Z}_{enc,t} \in \mathbb{R}^{d_r}$. We then employ an adaptive router to compute selection weights $w \in \mathbb{R}^{C}$. To facilitate exploration and prevent premature convergence to a suboptimal bottleneck, we inject input-dependent Gaussian noise into the routing mechanism. The selection mechanism is formulated as $w = \text{Softplus}(\bar{z} W_{router} + (\bar{z} W_{noise}) \odot \epsilon)$,
where $W_{router} \in \mathbb{R}^{d_{r} \times C}$ and $W_{noise} \in \mathbb{R}^{d_{r} \times C}$ are learnable weight matrices, $\epsilon \sim \mathcal{N}(0, I)$ is a standard Gaussian noise vector, and $\odot$ denotes the element-wise product. This noise term acts as a stochastic regularizer that maintains routing diversity. By perturbing the routing logits, the model is forced to explore diverse compression resolutions during the initial training phase, ensuring that all branches in the pool are effectively optimized and preventing mode collapse. As demonstrated by the ablation study in Appendix~\ref{sec:app_noise}, this noise-driven exploration significantly enhances \textit{branch utilization diversity} and improves generalization compared to deterministic routing.

\paragraph{Feature Integration.}
We employ a top-$k$ selection strategy to balance efficiency and representational capacity. Let $K$ be the set of indices corresponding to the $k$ largest values in $w$. The adaptive feature representation is obtained by a weighted fusion of the upsampled features from the selected branches:
\begin{equation}
    \mathbf{Z}_{aafi} = \sum_{i \in K} \frac{w_{i}}{\sum_{j \in K} w_{j}} \cdot \text{UpNet}_{i}(\mathbf{Z}_{i})
\end{equation}
where $\text{UpNet}_{i}$ maps the compressed features $\mathbf{Z}_{i}$ back to the original dimension $d_{r}$. The adaptive feature representation $\mathbf{Z}_{aafi}$ is subsequently fused with the shared encoder features $\mathbf{Z}_{enc}$. The combined representation is then projected via a fully connected layer to form the final acoustic context $H_{acous}$, following the integration process described in Section~\ref{sec:overview}.

\subsection{Mel-Band Adversarial Perturbation}
\label{sec:mel-bap}


\paragraph{Motivation: Mitigating the Robustness-Accuracy Trade-off.}
State-of-the-art deepfake detectors often exhibit a pathological reliance on imperceptible high-frequency artifacts as discriminative cues~\cite{zhang2025fsat}. While this yields exceptional performance on specific benchmarks, it renders models highly vulnerable to real-world distortions—such as lossy compression or environmental noise—that inevitably corrupt these fragile features. Although standard adversarial training or data augmentation can bolster robustness, they frequently incur a significant performance penalty on clean data, a dilemma recognized as the \textit{robustness-accuracy trade-off}~\cite{zhang2019theoretically}. We posit that this degradation stems from the nature of global perturbations, which often act as an indiscriminate "blunt instrument" that obscures essential semantic content and compromises the global spectro-temporal integrity of the audio.

To bridge this gap, we propose the Mel-Band Adversarial Perturbation (Mel-BAP) strategy. Unlike conventional global attacks, Mel-BAP applies \textit{band-selective perturbations} specifically targeting the mel-frequency regions where synthetic artifacts are most salient. This approach serves as a form of \textit{targeted regularization}: by smoothing the decision boundary within sensitive feature subspaces, it prevents the model from exploiting ``shortcuts'' based on unstable high-frequency patterns. Consequently, Mel-BAP incentivizes the model to shift its focus toward \textit{intrinsic, invariant representations} that characterize the underlying acoustic consistency. By constraining perturbations to localized bands, Mel-BAP effectively fortifies the model against both malicious adversarial attacks and heterogeneous real-world distortions without sacrificing the fidelity of global acoustic structures or clean-data accuracy.

\paragraph{Mechanism.}
Mel-BAP operates directly in the mel-spectrogram domain to align with the perceptual scaling of the Whisper encoder.
Let $\mathbf{M} \in \mathbb{R}^{T \times F}$ denote the input mel-spectrogram derived from audio $\mathbf{x}$, where $T$ is the number of time frames and $F=80$ is the number of mel-frequency bins. We define a binary mask $\mathbf{D} \in \{0, 1\}^{T \times F}$ to restrict perturbations to a target mel-band $[m_l, m_u]$:
\begin{equation}
\mathbf{D}_{t,f} = 
\begin{cases} 
1 & \text{if } f \in [m_l, m_u], \\
0 & \text{otherwise},
\end{cases}
\end{equation}

The adversarial perturbation $\boldsymbol{\delta}_m$ is optimized to maximize the detection loss $\mathcal{L}$ for the target model $\mathcal{F}_\theta$ within the constrained band. The perturbed input $\mathbf{M}'$ is computed as $\mathbf{M}' = \mathbf{M} + \mathbf{D} \odot \boldsymbol{\delta}_m$.
We solve the maximization problem $\max_{\boldsymbol{\delta}_m} \mathcal{L}(\mathcal{F}_\theta(\mathbf{M} + \mathbf{D} \odot \boldsymbol{\delta}_m), y)$ subject to an $L_{\infty}$-norm constraint $\|\boldsymbol{\delta}_m\|_\infty \leq \epsilon$ via projected gradient descent (PGD) \cite{madry2018towards}. 
Unlike standard adversarial training that often requires expensive parameter sweeps, we adopt a fixed configuration based on empirical stability:
\begin{equation}
\begin{aligned}
\boldsymbol{\delta}_m^{(k+1)} = \Pi_{\epsilon} \Bigl( & \boldsymbol{\delta}_m^{(k)} + \alpha \cdot \text{sign} \bigl(
 \nabla_{\boldsymbol{\delta}_m} \mathcal{L}(\mathcal{F}_\theta(\mathbf{M} + \mathbf{D} \odot \boldsymbol{\delta}_m^{(k)}), y) \bigr) \Bigr)
\end{aligned}
\end{equation}
Specifically, we set the perturbation budget $\epsilon=0.01$ (relative to the normalized spectral magnitude), the step size $\alpha=0.002$, and the number of iterations $K=5$. $\Pi_{\epsilon}$ projects the perturbation onto the $\epsilon$-ball, and $\odot$ denotes element-wise multiplication. A detailed ablation study on the accuracy-cost trade-off associated with the number of iterations $K$ is provided in Appendix~\ref{app:pgd_ablation}.


Mel-BAP shares the high-frequency regularization motivation of Frequency-Selective Adversarial Training (F-SAT, Zhang et al. \cite{zhang2025fsat}), while adapting the perturbation procedure to the Whisper pipeline:
\textit{1) Perceptual Alignment and Forensic Priors:} While F-SAT perturbs selected frequency bands in the STFT magnitude spectrum and reconstructs the waveform, Mel-BAP applies perturbations directly in Whisper's log-Mel input space. This aligns perturbation optimization with the representation consumed by the encoder. Mel-BAP targets high-frequency components to discourage reliance on fragile detection cues. Mel-BAP uses a fixed 4-8 kHz target band rather than random or sweeping band selection. We provide a frequency-band ablation in Table~\ref{tab:melbap_band_ablation} to compare this constraint with full-band adversarial training.
\textit{2) End-to-End Computational Efficiency:} By modifying the log-Mel input directly, Mel-BAP avoids the waveform reconstruction and subsequent feature extraction required by the F-SAT pipeline, keeping perturbation optimization within the detector's input representation.
\textit{3) Zero Inference Cost:} Mel-BAP operates exclusively during training. It therefore adds no perturbation-generation overhead during inference.


\section{Experimental Evaluation}
\label{sec: Experimental settings}

\paragraph{Experimental Settings.} 
Unless otherwise stated, our experiments primarily use the Bilingual Dataset (BD) from SpeechFake\footnote{Available from: \url{https://www.modelscope.cn/datasets/inclusionAI/SPEECHFAKE}} \cite{huang2025speechfake}, which contains high-fidelity deepfakes in English (EN) and Chinese (CN). The dataset includes samples generated using 30 open-source tools (e.g., TorToiSe, VITS, FastSpeech2) and 10 commercial APIs (e.g., Azure TTS, Google Wavenet), spanning text-to-speech (TTS), voice conversion (VC), and neural vocoder (NV) architectures. Based on the subset splits used in our evaluation, the BD dataset comprises over 1.19 million English samples and over 0.88 million Chinese samples, totaling more than 2 million high-fidelity deepfake and genuine utterances. We use the official SpeechFake BD train, dev, and test subsets, containing 75,708/12,618/37,854 real and 633,354/105,544/315,222 fake utterances, respectively. No samples are reassigned across the released subsets. An unseen test set (BD-UT) contains over 85,000 test samples, with all fake samples generated exclusively using commercial APIs unseen during training. Detailed dataset statistics are provided in Table~\ref{tab:bd_subsets} in Appendix~\ref{sec:Dataset Statistics}.

\begin{table}[t]
    \centering
    \captionsetup{skip=7pt}
    \caption{Performance Comparison of Adaptive Routing (top-$k$) in the AAFI Module versus Whisper base. EER (\%) results (lower is better).}
    \label{tab:expected_adaptive_routing}
    \footnotesize
    \renewcommand{\arraystretch}{1.15}
    \setlength{\tabcolsep}{0pt} 
    \begin{tabular*}{\linewidth}{@{\extracolsep{\fill}} c ccccc}
        \toprule
        \multirow{2}{*}{\textbf{Data Group}} & \multicolumn{4}{c}{\textbf{top-$k$ number}} & \multirowcell{2}{\textbf{Whisper-base}} \\
        \cmidrule{2-5}
        & \textbf{top-2} & \textbf{top-3} & \textbf{top-4} & \textbf{top-5} & \\
        \midrule
        TTS & 3.91 & 0.85 & \textbf{0.83} & 0.99 & 1.18 \\
        VC  & 4.85 & 1.88 & \textbf{1.67} & 2.11 & 3.32 \\
        NV  & 8.44 & \textbf{3.44} & 3.50 & 4.34 & 6.07 \\
        \midrule 
        Average & 5.73 & 2.06 & \textbf{2.00} & 2.48 & 3.52 \\
        \bottomrule
    \end{tabular*}
\end{table}

All experiments were conducted on two NVIDIA A6000 GPUs. We optimized the model using AdamW \cite{zhang2019theoretically} with a learning rate of $1\times10^{-4}$ and a weight decay of $1\times10^{-5}$. To address class imbalance, we employed a weighted cross-entropy loss with weights of $1.0$ for genuine samples and $0.6$ for spoofed samples. Training used a batch size of 32 for up to 10 epochs, with early stopping triggered if the validation EER did not improve for five consecutive epochs. Inputs were 16~kHz raw waveforms. We evaluated performance using the widely adopted EER metric. Our base models were Whisper-base (74M parameters) and Whisper-small (244M parameters).

To apply Mel-BAP, the audio input $\mathbf{x}$ is first converted into a log-Mel spectrogram $\mathbf{M} \in \mathbb{R}^{T \times F}$ (where $F=80$ is the number of mel bins) using Whisper’s default preprocessing pipeline (25 ms window, 10 ms hop, 80 mel bins, log-compression). A target Mel-bin range $[m_l, m_u]$ is then selected based on artifact analysis (e.g., $m_l=60$ and $m_u=80$). Within this band, adversarial perturbations $\boldsymbol{\delta}_m$ are generated using five PGD iterations and added to obtain the perturbed spectrogram $\mathbf{M}' = \mathbf{M} + \mathbf{D} \odot \boldsymbol{\delta}_m$, which is subsequently fed into Whisper’s encoder for prediction.

\subsection{AAFI Ablation and Routing Analysis}

We conduct targeted experiments on the SpeechFake dataset to validate the core functionality and design choices of the AAFI module. These experiments focus on two key aspects: (1) whether AAFI dynamically adapts to input characteristics as intended, and (2) how to select the optimal value of $k$ for top-$k$ routing. All AAFI-only variants are trained on the BD training set under identical conditions. We evaluate them on the three SpeechFake BD validation subsets (TTS, VC, and NV). For each sample, we record the compression dimensions (16/32/64/128/256) selected by AAFI's adaptive router. To reduce computational cost and ensure a fair comparison, we freeze all encoder layers across all evaluated configurations (including the Whisper-base baseline).

Table~\ref{tab:expected_adaptive_routing} compares the performance of the adaptive routing strategy within the AAFI module under different top-$k$ configurations against the base model without AAFI. Incorporating AAFI with a sufficient routing capacity ($k \ge 3$) results in a clear improvement in EER across all validation groups. Conversely, excessively restricting the feature integration to top-2 degrades performance below the baseline (5.73\% vs. 3.52\% in average EER), indicating that sufficient diversity in compression resolutions is essential. Although top-4 achieves the best performance (2.00\% in average EER), the difference between top-3 and top-4 is minimal (0.85\% vs. 0.83\% for TTS, 1.88\% vs. 1.67\% for VC, 3.44\% vs. 3.50\% for NV, and 2.06\% vs. 2.00\% on average), suggesting that top-3 already provides nearly optimal performance. Extending to top-5 slightly reduces performance, likely due to the inclusion of redundant information, while increasing computational cost. Overall, top-3 offers the best trade-off between efficiency and accuracy and is the most practical choice for deployment. Furthermore, a stratified analysis shows that routing distributions vary across spoofing conditions: NV samples assign 59\% of their selections to the wider 128/256-d branches, whereas TTS samples assign 64\% to the narrower 16/32-d branches. The observed distributions are consistent with condition-dependent multi-resolution selection; full results are provided in Appendix~\ref{app:routing_distribution}.

\subsection{Mel-BAP Ablation Study}

\paragraph{Module Ablation.}

We evaluate Mel-BAP's efficacy by comparing \textit{1) Whisper base}, \textit{2) Base+Mel-BAP}, and \textit{3) AAFI+Mel-BAP} across SpeechFake subsets. Table~\ref{tab:ablation_Mel-BAP} shows that Mel-BAP consistently reduces EER (e.g., from 1.19\% to 0.91\% on BD-UT), demonstrating its ability to stabilize predictions against unseen spoofing patterns. Integrating AAFI further lowers EER to 0.53\%, confirming that adaptive feature integration complements targeted perturbations by selecting optimal representations for different inputs. 

\paragraph{Frequency-Band Ablation.}
To examine the effect of the frequency-band constraint, we compare high-, full-, and low-band perturbations using Whisper-base without AAFI or standard augmentation. All variants use the same perturbation budget ($\epsilon=0.01$, $\alpha=0.002$, $K=5$). Table~\ref{tab:melbap_band_ablation} shows that Mel-BAP achieves an EER of 0.91\%, compared with 1.15\% for full-band PGD and 1.28\% for low-band PGD.

\subsection{Robustness under Real-World Distortions}

To assess real-world resilience, we subjected the BD-UT set to dynamic attacks, including background noise (NoiseX-92 \cite{varga1993assessment}), speed adjustments (0.5$\times$--1.5$\times$), and volume fluctuations ($\pm$15 dBFS). The results are reported in Table~\ref{tab:ablation_real-world}. 
\textit{First,} Mel-BAP significantly enhances robustness across all tested distortions. For both Whisper-base and Whisper-small, it consistently reduces EER in both clean and perturbed conditions (e.g., lowering EER under dynamic attacks from 3.57\% to 2.44\% for Whisper-base).
\textit{Second, and most critically,} Mel-BAP demonstrates a decisive advantage over standard global audio augmentation. As shown in the ``Global Aug'' columns, \textit{even when the model is trained with the identical types of perturbations used in our evaluation (i.e., the same noise, speed, and volume variations)}, simply applying them as indiscriminate global distortions during training significantly degrades performance. This leads to substantial performance degradation even on clean data (e.g., EER rising from 1.19\% to 2.33\% for Whisper-base). This result highlights a classic robustness-accuracy trade-off, in which global perturbations may inadvertently destroy essential semantic and structural cues. In sharp contrast, Mel-BAP’s band-selective strategy functions as a \textit{targeted regularizer}. By forcing the model to ignore fragile high-frequency artifacts while preserving the overall acoustic structure, it encourages the extraction of intrinsic, invariant features rather than overfitting to noise patterns.
\textit{Finally,} while larger models (Whisper-small) provide stronger baselines, the relative improvement from Mel-BAP is more pronounced in Whisper-base. This suggests that specialized mechanisms like Mel-BAP are crucial for preventing shortcut learning and sustaining generalization in non-stationary, real-world scenarios.

Beyond standard real-world distortions, we evaluate the adversarial robustness of our framework against strong test-time adaptive attacks (including standard PGD, CW, and frequency-aware Adaptive PGD). Detailed evaluation protocols and results are provided in Appendix~\ref{app:adaptive_attack}. 

\begin{table}[t]
    \centering
    \captionsetup{skip=7pt}
    \renewcommand{\arraystretch}{1.15}
    \footnotesize 
    \setlength{\tabcolsep}{0pt} 
    \caption{Ablation study on the effectiveness of Mel-BAP across SpeechFake BD subsets. Lower EER (\%) indicates better performance.}
    \label{tab:ablation_Mel-BAP}
    \begin{tabular*}{\linewidth}{@{\extracolsep{\fill}} c cccc}
        \toprule
        \textbf{Model} & \textbf{BD-EN} & \textbf{BD-CN} & \textbf{BD-UT} & \textbf{Average}\\
        \midrule
        w/o Mel-BAP     & 1.94 & 1.68 & 1.19 & 1.60 \\
        Mel-BAP Only    & 1.35 & 1.20 & 0.91 & 1.14 \\
        AAFI + Mel-BAP  & \textbf{1.01} & \textbf{0.82} & \textbf{0.53} & \textbf{0.79} \\
        \bottomrule
    \end{tabular*}
\end{table}

\begin{table}[t]
    \centering
    \captionsetup{skip=7pt}
    \renewcommand{\arraystretch}{1.15}
    \setlength{\tabcolsep}{0pt}
    \caption{Ablation of adversarial frequency band constraints on the BD-UT dataset (Clean EER).}
    \label{tab:melbap_band_ablation}
    \begin{tabular*}{\linewidth}{@{\extracolsep{\fill}} cccc}
        \toprule
        \textbf{Strategy} & \makecell{\textbf{Frequency}\\\textbf{Band}} & \makecell{\textbf{Mel-bin}\\\textbf{Indices}} & \makecell{\textbf{Clean}\\\textbf{EER (\%)}} \\
        \midrule
        (a) \textbf{Mel-BAP (Ours)} & 4--8 kHz & [60, 80] & \textbf{0.91} \\
        (b) Full-band PGD & 0--8 kHz & [0, 80] & 1.15 \\
        (c) Low-band PGD & 0--4 kHz & [0, 60] & 1.28 \\
        \bottomrule
    \end{tabular*}
\end{table}

\begin{table}[t]
    \centering
    \captionsetup{skip=7pt}
    \renewcommand{\arraystretch}{1.15}
    \small 
    \caption{Robustness comparison against standard augmentation under real-world distortions. EER (\%) is reported on BD-UT.}
    \label{tab:ablation_real-world}
    \setlength{\tabcolsep}{0pt}
    \begin{tabular*}{\linewidth}{@{\extracolsep{\fill}} c cccccc}
        \toprule
        \multirow{2}{*}{\textbf{Model}} & \multicolumn{2}{c}{\textbf{w/o Mel-BAP}} & \multicolumn{2}{c}{\textbf{with Mel-BAP}} & \multicolumn{2}{c}{\textbf{Global Aug}} \\
        \cmidrule{2-3} \cmidrule{4-5} \cmidrule{6-7}
        & Clean & Attack & Clean & Attack & Clean & Attack \\
        \midrule
        Base  & 1.19 & 3.57 & \textbf{0.91} & \textbf{2.44} & 2.33 & 4.68 \\
        Small & 0.95 & 3.18 & \textbf{0.87} & \textbf{2.25} & 1.65 & 4.01 \\
        \bottomrule
    \end{tabular*}

\end{table}

\begin{table*}[t]
\centering
\captionsetup{skip=7pt}
\setlength{\tabcolsep}{3pt}
\renewcommand{\arraystretch}{1.2}
\tiny
\caption{Comprehensive Zero-shot EER (\%) results across 29 systems. All models are evaluated without fine-tuning. The Average is recalculated based on five key benchmarks: In-the-Wild, ASV2019, ASV2021 LA, FakeOrReal, and ADD 2022 Track 1. Values in bold indicate the best results per column. Lower EER indicates superior performance. }
\label{tab:comprehensive_zero_shot}
\resizebox{\linewidth}{!}{
\begin{tabular}{@{}lccccccc@{}}
\toprule
\textbf{Model ID} & \textbf{\# Params} & \textbf{ADD 2022} (T1) & \textbf{FakeOrReal} & \textbf{In-the-Wild} & \textbf{ASV2019} & \textbf{ASV2021 LA} & \textbf{Average} \\
\midrule
\multicolumn{8}{l}{\textit{Literature and Commercial SOTA (Speech-DF-Arena 2025)}} \\

\midrule
Resemble-Detect-3B-Omni~\cite{dowerah2025df} & 3,000 M & 9.96 & 2.34 & 1.35 & 2.53 & 3.02 & 3.84 \\
DLMSL-SpeakSure-v0.1~\cite{dowerah2025df}    & 658.6 M & 15.14 & 0.27 & 1.28 & \textbf{0.04} & \textbf{0.08} & 3.36 \\
Whispeak~\cite{dowerah2025df}                & 98.9 M  & 11.94 & 1.02 & 1.27 & 0.39 & 3.58 & 3.64 \\
DF\_Arena\_1B\_V\_1~\cite{kulkarni2025df}         & 1,000 M & 22.21 & 2.92 & 0.91 & 1.14 & 4.66 & 6.37 \\
Momenta~\cite{dowerah2025df}                   & 350 M & 25.47 & 2.87 & 4.63 & 1.21 & 4.88 & 7.81 \\
Syntra Detector~\cite{dowerah2025df}           & 584 M & 23.58 & 0.53 & 3.98 & 1.48 & 14.05& 8.73 \\
DF\_Arena\_500M\_V\_1~\cite{dowerah2025df}        & 500 M & 23.98 & 2.30 & 1.76 & 1.09 & 4.23 & 6.67 \\
MoLEx~\cite{dowerah2025df}                     & 376.4 M & 31.94 & \textbf{0.18} & \textbf{0.03} & 0.28 & 6.33 & 7.75 \\
Resemble Detect~\cite{dowerah2025df}           & 2,112 M & 28.22 & 1.37 & 3.95 & 1.32 & 1.65 & 7.30 \\
DF\_Arena\_100M\_V\_1~\cite{dowerah2025df}      & 100 M & 27.06 & 7.42 & 2.22 & 1.54 & 7.61 & 9.17 \\
DF\_Arena\_100M\_V\_0~\cite{dowerah2025df}      & 100 M & 31.96 & 6.14 & 4.29 & 4.16 & 10.06& 11.32\\
XLSR + SLS~\cite{zhang2024deepfake}                & 340 M & 33.95 & 5.08 & 7.46 & 0.23 & 2.87 & 9.92 \\
TCM~\cite{truong2024temporal2}                       & 319 M & 37.40 & 10.69& 7.79 & 0.19 & 3.00 & 11.81\\
BiCrossMamba-ST~\cite{elkheir2025bicrossmamba2}           & 318.2 M & 30.44 & 6.85 & 7.94 & 0.71 & 3.83 & 9.95 \\
Nes2NetX~\cite{liu2025nes2net}                  & 317.9 M & 34.47 & 6.32 & 7.75 & 0.12 & 2.17 & 10.17\\
Wav2Vec2 AASIST~\cite{tak2022automatic}           & 317.8 M & 31.05 & 7.47 & 11.20& 0.22 & 0.82 & 10.15\\
XLSR Mamba~\cite{xiao2025xlsr}                & 319 M & 34.23 & 6.71 & 6.71 & 0.42 & 0.93 & 9.80 \\
Whisper Mesonet~\cite{kawa2023improved}           & 7.6 M   & 38.38 & 47.75& 26.73& 5.83 & 15.82& 26.90\\
Wav2Vec2 ECAPA~\cite{kulkarni2024exploring}            & 324 M & 46.44 & 62.32& 30.70& 29.70& 26.61& 39.15\\
AASIST~\cite{jung2022aasist2}                    & 0.3 M   & 51.06 & 21.64& 43.01& 0.83 & 11.46& 25.60\\
WavLM ECAPA~\cite{kulkarni2024exploring}               & 102 M & 44.17 & 23.37& 34.65& 0.76 & 6.68 & 21.92\\
RawGatST~\cite{tak2021spectro}                  & 0.4 M   & 50.00 & 53.09& 52.54& 1.06 & 10.25& 33.39\\
RawTFNet~\cite{xiao2025rawtfnet}                  & 0.2 M   & 43.85 & 36.57& 38.73& 1.89 & 5.04 & 25.22\\
Hubert ECAPA~\cite{kulkarni2024exploring}              & 102 M & 47.66 & 46.22& 38.66& 1.06 & 12.55& 29.23\\
Rawnet2~\cite{tak2021anti}                   & 17.6 M  & 49.36 & 65.68& 49.19& 33.04& 40.07& 47.47\\
\midrule
\multicolumn{8}{l}{\textit{Our Proposed Models}} \\
\midrule
Whisper-base w/o (AAFI+Mel-BAP) & 74 M & 3.11 & 0.91 & 5.53 & 1.94 & 2.15 & 2.73 \\
Whisper-base + (AAFI+Mel-BAP)   & 80 M & 2.16 & 0.55 & 4.78 & 1.19 & 1.30 & 2.00 \\
Whisper-small w/o (AAFI+Mel-BAP) & 244 M & 2.94 & 0.73 & 4.36 & 1.10 & 1.77 & 2.18 \\
Whisper-small + (AAFI+Mel-BAP)  & 252 M & \textbf{1.87} & 0.49 & 3.14 & 0.93 & 1.12 & \textbf{1.51} \\
\bottomrule
\end{tabular}
}
\end{table*}

\subsection{Comparison with State-of-The-Art Methods}

\paragraph{Fairness of Comparison and Zero-shot Protocol.}
A strictly fair comparison with state-of-the-art (SOTA) systems is challenging due to the diversity of proprietary training data used in commercial models (e.g., DLMSL, Resemble) and varying baselines.
Therefore, our evaluation focuses on \textit{zero-shot generalization potential} in real-world scenarios rather than algorithmic superiority under controlled training conditions. Results for the 25 external systems in Table~\ref{tab:comprehensive_zero_shot} are taken from Speech-DF-Arena 2025~\cite{dowerah2025df}. Because these systems use heterogeneous or undisclosed training protocols, the table provides a contextual rather than strictly controlled comparison. Our four Whisper-based models were trained on the same official SpeechFake BD splits and evaluated under the same zero-shot protocol. The term ``zero-shot'' here refers to evaluation on datasets (e.g., ADD 2022, In-the-Wild) not used during training.

To ensure a comprehensive evaluation across diverse and realistic scenarios, we evaluated our models on FakeOrReal~\cite{reimao2019dataset} and its second version FakeOrReal-norm, In-the-Wild~\cite{muller2022does}, the Audio Deepfake Detection Challenge (ADD 2022) track 1 evaluation set~\cite{yi2022add}, ASVspoofing 2019~\cite{todisco2019asvspoof} and ASVspoofing 2021 LA~\cite{yamagishi2021asvspoof}. FakeOrReal and In-the-Wild are English-language datasets that are widely used in the deepfake detection community to assess a detector’s generalization performance. ADD 2022 is a Mandarin-language challenge dataset, and its deepfake speech samples were collected from the generation task of the same challenge. This selection covers the full spectrum of modern deepfake threats, ranging from high-fidelity lab-generated samples to unconstrained real-world recordings, thereby surpassing the limitations of traditional controlled benchmarks. In our study, we used only the audio portion, which includes samples in more than 50 languages. We used only the test partitions of the selected test datasets; their training portions, if available, were excluded entirely from both the model training and testing.

Table~\ref{tab:comprehensive_zero_shot} presents a \textit{comprehensive zero-shot evaluation} against 25 state-of-the-art systems and classic baselines. The results reveal several critical insights into the current landscape of deepfake detection. \textit{First,} while certain specialized models exhibit exceptional performance on specific benchmarks, they often suffer from a ``generalization collapse'' in adverse environments. For instance, MoLEx achieves near-perfect results on \textit{FakeOrReal} ($0.18\%$) and \textit{In-the-Wild} ($0.03\%$), yet its performance catastrophically degrades to $31.94\%$ on ADD 2022 Track 1. Similarly, the commercial leader DLMSL-SpeakSure-v0.1 dominates the clean \textit{ASV2019 and ASV2021 LA} benchmarks ($<0.1\%$) but struggles with the complex distortions in ADD 2022 ($15.14\%$). This discrepancy suggests that these models may rely on ``shortcuts'', namely fragile high-frequency artifacts that are easily obliterated by real-world noise or lossy compression. \textit{Second,} our proposed \textit{Whisper-small + AAFI + Mel-BAP} model (252M parameters) emerges as the most robust generalist, delivering the lowest overall \textit{average EER of $1.51\%$}. Most notably, it achieves a SOTA EER of \textit{$1.87\%$} on ADD 2022 Track 1, outperforming the 3-billion-parameter Resemble-Detect-3B-Omni ($9.96\%$) by a factor of five despite using approximately $11.9\times$ fewer parameters. This demonstrates that Mel-BAP effectively incentivizes the model to capture intrinsic acoustic consistency rather than overfitting to dataset-specific noise patterns. \textit{Third,} the proposed approach maintains a superior efficiency-performance Pareto frontier. With only 252M parameters, it achieves a 76.3\% lower average EER than the DF\_Arena\_1B\_V\_1 with 1 billion parameters ($1.51\%$ vs. $6.37\%$) and is significantly more accurate than massive commercial models. The integration of the AAFI + Mel-BAP modules allows the detector to generalize to novel forgery techniques and dynamic attack scenarios without fine-tuning.


\section{Conclusion}

This paper presents a robust and scalable framework for generalized ADD by integrating AAFI and Mel-BAP into a Whisper-based architecture. Our approach effectively bridges the gap between representation fidelity and adversarial robustness. By leveraging a stochastic compression pool and targeted mel-domain regularization, the framework achieves superior zero-shot generalization across diverse languages and synthesis domains. Experimental results confirm that our model establishes a new efficiency-performance Pareto frontier for ADD, offering a modular ``plug-in'' solution that can be seamlessly integrated into existing security pipelines. Future work will focus on further refining adaptive mechanisms to support scalable real-time detection in non-stationary acoustic environments.

\section*{Acknowledgements}
This research was supported in part by the Self-initiated Project of Yangtze Delta Region Institute of Tsinghua University, Zhejiang No. 140104370.
All authors are grateful to the anonymous referees for their valuable comments that led to considerable improvements to this article.

\bibliographystyle{unsrt}
\bibliography{reference_cr}

\newpage
\appendix

\label{sec:appendix}

\section{Ablation Study on Unseen Languages}
\label{sec:cross-lingual}

To assess the cross-lingual generalization ability of our proposed model in the ADD task, we conduct an ablation study on the SpeechFake Multilingual dataset (MD)~\cite{huang2025speechfake}. The MD covers a total of 46 languages, including 9 primary languages---English (en), Chinese (zh), Spanish (es), French (fr), Hindi (hi), Japanese (ja), Korean (ko), Persian (fa), and Italian (it)---together with 37 additional languages. Real speech samples in MD are sourced from the CommonVoice corpus~\cite{ardila2020common}, ensuring broad multilingual coverage.  

For this experiment, we evaluate the \textbf{Whisper-base} model with both AAFI and Mel-BAP. The training data are drawn exclusively from the SpeechFake BD training subset, which contains only Chinese and English speech, thereby simulating a limited bilingual training scenario. During testing, we focus on a representative set of unseen languages, including Spanish (es), French (fr), Japanese (ja), Italian (it), and Korean (ko). For each language, we randomly sample 10,000 utterances from the MD for evaluation. Notably, the samples from the MD share nearly no overlap with BD. Generalization to MD is therefore more challenging than to BD validation subsets.  This setup allows us to rigorously examine the model’s robustness to cross-lingual transfer, particularly under conditions where the training and testing languages are entirely disjoint.

Table~\ref{tab:crosslingual_results} reports the cross-lingual evaluation results on five unseen languages from the SpeechFake MD dataset. Several key observations can be made:

\begin{itemize}[leftmargin=*, labelsep=0.5em]
\item 
First, across all languages, the Whisper-base model equipped with AAFI and Mel-BAP consistently outperforms the baseline without these components, achieving lower EER values. This confirms that the proposed strategies not only improve detection accuracy in in-domain conditions but also enhance robustness under cross-lingual settings where the model encounters unseen languages. 
\item 
Second, the degree of improvement varies by language. The most notable gains are observed for Spanish and French, where the integration of AAFI and Mel-BAP reduces the EER by approximately 1.5\% and 1.2\%, respectively. This suggests that these modules effectively capture language-invariant spoofing cues, enabling better generalization across linguistic boundaries. For other languages such as Japanese, Italian, and Korean, the improvements remain consistent but slightly smaller, indicating that language-specific acoustic and phonetic properties may affect the transferability of spoofing detection features. 
\item 
Finally, the results reinforce the effectiveness of Whisper as a baseline model for ADD. Thanks to Whisper's original powerful multilingual transcription ability, when combined with AAFI and Mel-BAP, the proposed model not only leverages its strong transcription and representation learning capabilities but also demonstrates resilience against cross-lingual distribution shifts.
\end{itemize}

\begin{table}[t]
    \centering
    \captionsetup{skip=7pt}
    \renewcommand{\arraystretch}{1.15}
    \footnotesize 
    \setlength{\tabcolsep}{0pt} 
    \caption{Cross-lingual evaluation results on unseen languages (SpeechFake MD). EER (\%) is compared between the baseline and our proposed method.}
    \label{tab:crosslingual_results}
    \begin{tabular*}{\linewidth}{@{\extracolsep{\fill}} c cc c @{}}
        \toprule
        \multirowcell{2}{\textbf{Language}} & \multicolumn{2}{c}{\textbf{Whisper-base EER (\%)}} & \multirowcell{2}{\textbf{Improv. $\downarrow$}}\\
        \cmidrule{2-3}
        & \textbf{w/o AAFI+BAP} & \textbf{w/ AAFI+BAP} & \\
        \midrule
        Spanish (es)  & 7.11 & \textbf{5.65} & 1.46 \\
        French (fr)   & 5.17 & \textbf{3.92} & 1.25 \\
        Japanese (ja) & 5.52 & \textbf{4.19} & 1.33 \\
        Italian (it)  & 4.66 & \textbf{3.91} & 0.75 \\
        Korean (ko)   & 4.50 & \textbf{4.02} & 0.48 \\
        \midrule
        \textbf{Average} & \textbf{5.39} & \textbf{4.34} & \textbf{1.05} \\
        \bottomrule
    \end{tabular*}
\end{table}


\section{Dataset Statistics}
\label{sec:Dataset Statistics}

CodecFake \cite{wu2024codecfake2} (shown in Table \ref{tab:codec_models_with_sub_id}), derived from the VCTK corpus \cite{yamagishi2019cstr}, focuses on neural codec-based synthesis. It contains over one million samples generated via various codec frameworks, providing a rigorous benchmark for detecting increasingly prevalent codec-based forgeries.

Table \ref{tab:bd_subsets} presents the detailed statistics of the SpeechFake Bilingual Dataset (BD) subsets used in our experiments.

\begin{table}[t]
\centering
\captionsetup{skip=7pt}
\renewcommand{\arraystretch}{1.2}
\scriptsize
\setlength{\tabcolsep}{0pt} 
\caption{Details of adopted neural audio codec models. $A-F$ represents different codec frameworks. $N_c$ represents the number of codebooks, SR represents audio sampling rate (in kHz), and BPS represents bits per second.}
\label{tab:codec_models_with_sub_id}
\begin{tabular*}{\linewidth}{@{\extracolsep{\fill}} c c c l c ccc @{}}
\toprule
\textbf{ID} & \textbf{Sub-ID} & \textbf{Model} & \textbf{Name} & \textbf{Training Data} & \bm{$N_c$} & \textbf{SR} & \textbf{BPS} \\
\midrule
A & A1 & SpeechTokenizer & 16k & Librispeech & 8 & 16 & 4 \\
\midrule
\multirow{3}{*}{B} & B1 & \multirow{3}{*}{HiFi-Codec} & hifi\_16k\_320d & LibriTTS & 4 & 16 & 2 \\
 & B2 & & hifi\_16k\_320d\_large\_uni & VCTK & 4 & 16 & 2 \\
 & B3 & & hifi\_24k\_320d & AISHELL & 4 & 24 & 3 \\
\midrule
C & C1 & AudioDec & 24k\_320d & Valentini & 8 & 24 & 6.4 \\
\midrule
\multirow{3}{*}{D} & D1 & \multirow{3}{*}{EnCodec} & 16k & Common Voice, DAPS & 12 & 16 & 6 \\
 & D2 & & 24k & MUSDB, VCTK & 32 & 24 & 24 \\
 & D3 & & 44k & AudioSet, Jamendo & 9 & 44.1 & 8 \\
\midrule
E & E1 & DAC & 24k & \makecell[c]{Common Voice, DAPS, Jamendo, \\ AudioSet, FSD50K} & 8 & 24 & 1.5 \\
\midrule
\multirow{6}{*}{F} & F1 & \multirow{6}{*}{FunCodec} & en\_libritts\_16k\_gr1nq32ds320 & \multirow{4}{*}{LibriTTS Subset} & 32 & 16 & 16 \\
 & F2 & & en\_libritts\_16k\_gr8nq32ds320 & & 32 & 16 & 16 \\
 & F3 & & en\_libritts\_16k\_nq32ds320 & & 32 & 16 & 16 \\
 & F4 & & en\_libritts\_16k\_nq32ds640 & & 32 & 16 & 8 \\
 & F5 & & zh\_en\_16k\_nq32ds320 & \multirow{2}{*}{\makecell[c]{In-house, 25k hrs \\ (en \& zh-cn)}} & 32 & 16 & 16 \\
 & F6 & & zh\_en\_16k\_nq32ds640 & & 32 & 16 & 8 \\
\bottomrule
\end{tabular*}
\end{table}

\begin{table}[htbp]
  \centering
  \captionsetup{skip=7pt}
  \renewcommand{\arraystretch}{1.15}
  \setlength{\tabcolsep}{0pt}
  \small 
  \caption{Statistics of SpeechFake Bilingual Dataset (BD) Subsets.}
  \label{tab:bd_subsets}
  \begin{tabular*}{\linewidth}{@{\extracolsep{\fill}} cccc}
    \toprule
    \multirowcell{2}{\textbf{Subset}} & \multirowcell{2}{\textbf{Content}} & \multicolumn{2}{c}{\textbf{Number of Samples}} \\
    \cmidrule{3-4}
                    &                  & \textbf{Real} & \textbf{Fake} \\
    \midrule
    BD-EN    & English-only     & 64,000  & 1,128,882 \\
    BD-CN    & Chinese-only     & 62,180  &   824,134 \\
    BD-UT    & Unseen Techniques & 37,854  &    50,000 \\
    BD-TTS   & TTS-generated    &126,180  & 1,014,107 \\
    BD-VC    & VC-generated     &126,180  &   535,205 \\
    BD-NV    & NV-generated     &126,180  &   403,704 \\
    \bottomrule
  \end{tabular*}
\end{table}


\begin{figure*}[t] 
    \centering
    \includegraphics[width=\textwidth]{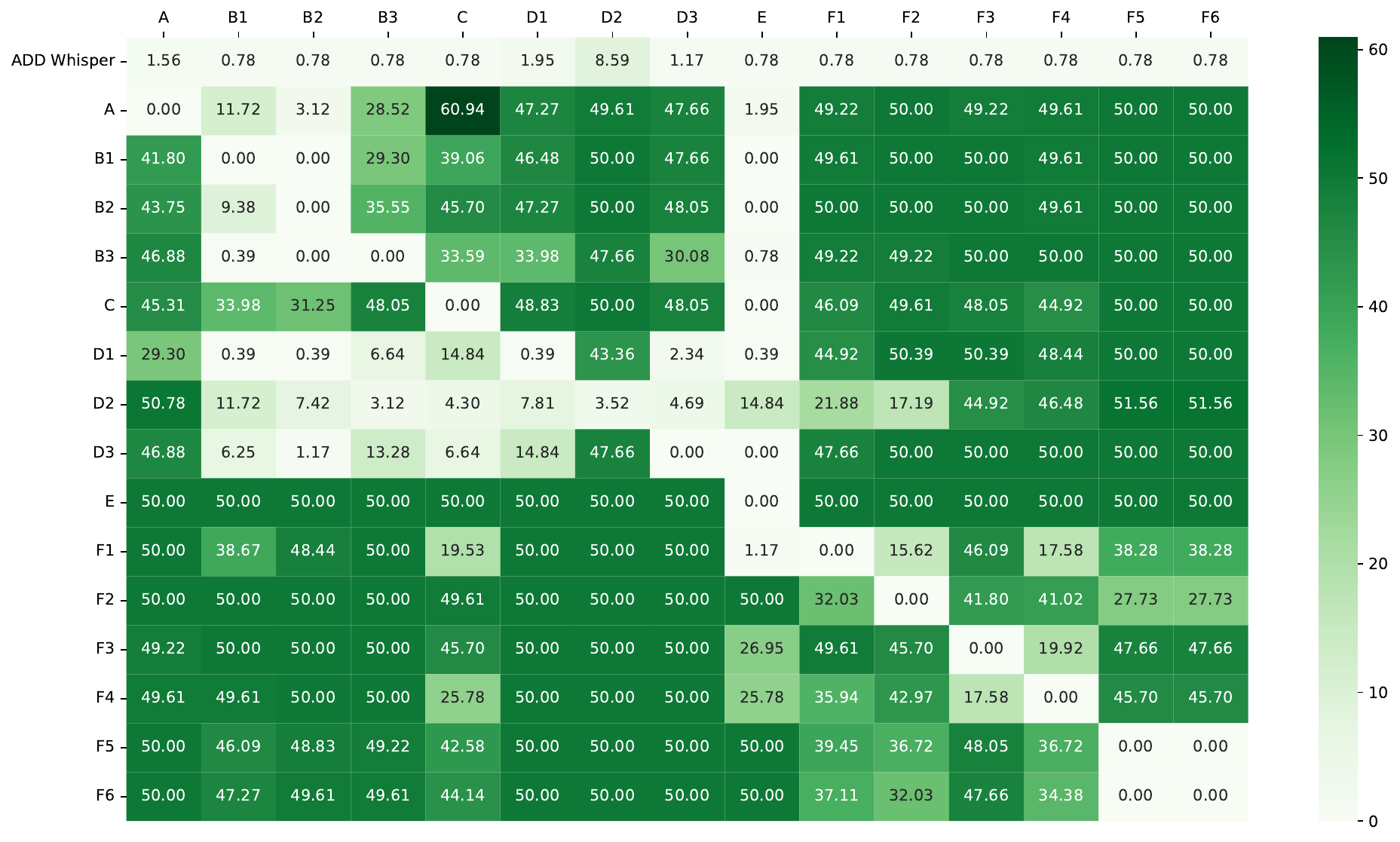} 
    \caption{Cross-testing results of anti-spoofing models on CodecFake test subsets. We report equal error rate (EER), where lower is better. Rows A-F6 are ADD whisper models trained on one training subset of CodecFake. Columns A-F6 indicate the specific CodecFake test subset evaluated.}
    \label{fig:cross test}
\end{figure*}

\section{Robustness and Generalization in Cross-Subset Testing Based on Whisper}
\label{sec:app_crosstest}

Cross-testing is essential to evaluate whether the model can generalize across diverse audio inputs derived from different types. This setting provides a robust assessment of the model’s adaptability and consistency under varying data distributions. In this ablation study, we used the CodecFake dataset for the experiment, where the dataset was split into training (80\%), validation (10\%), and test (10\%) sets, independently. All cross-testing experiments were conducted using the Whisper-base model (labelled ``ADD whisper'') on the CodecFake dataset, with results illustrated in Figure~\ref{fig:cross test}. The figure presents the Equal Error Rate (EER) across different training and testing subsets of the CodecFake dataset. Each row corresponds to a distinct training subset, while each column represents a testing subset, forming a matrix-like visualization that comprehensively reflects performance across subset combinations. The key findings are as follows:

The model trained on the full CodecFake dataset (row 1) achieves consistently lower EER values across nearly all test subsets, demonstrating strong generalization across diverse spoofing conditions. This advantage stems from the model’s exposure to a wide distribution of codec-induced variations, which enables it to capture richer spoofing patterns and reduces the risk of overfitting to narrow conditions. In contrast, models trained on individual subsets (rows A--F6) tend to specialize in the corresponding codec characteristics and thus achieve relatively low EERs on their in-distribution counterparts (e.g., row A $\rightarrow$ column A or row B1 $\rightarrow$ columns B1--B3). However, their performance degrades markedly under out-of-distribution testing, where EER values increase due to the lack of exposure to unseen codec artifacts. This contrast highlights a key finding: equipping the proposed model with more comprehensive knowledge of spoofing patterns yields consistently lower error rates and superior robustness. 

Furthermore, the effectiveness of Whisper in this setting underscores its validity as a strong baseline for audio deepfake detection. Owing to its large-scale pre-training on diverse multilingual and multi-domain speech data, Whisper provides robust acoustic and semantic representations that transfer effectively to anti-spoofing tasks. When fine-tuning with structured prompts, as in ADD Whisper, the model not only preserves its transcription capability but also learns to associate subtle semantic–acoustic inconsistencies with spoofing cues. This dual capability makes Whisper particularly well-suited for evaluating and benchmarking ADD methods. Taken together, our results confirm that combining a powerful pre-trained backbone with comprehensive training data is critical for building reliable and generalizable deepfake detection systems.

\section{Implementation Details of AAFI}
\label{app:aafi_details}

\subsection{Architectural Specifications and Complexity}
To ensure that the performance gains stem from adaptive feature selection rather than increased model capacity, the AAFI module is designed to be lightweight.
\begin{itemize}
    \item \textbf{Dimensionality ($d_r$):} The input dimension $d_r$ corresponds strictly to the hidden size of the backbone encoder: $d_r=512$ for Whisper-base and $d_r=768$ for Whisper-small.
    \item \textbf{Projection Layers:} Both $\text{DownNet}_{i}$ and $\text{UpNet}_{i}$ consist of a single linear transformation (fully connected layer) without bias, which intrinsically imposes the information bottleneck by projecting the features to the reduced latent dimension.
    \item \textbf{Parameter Overhead:} The total number of parameters introduced purely by the heterogeneous compression pool (with latent dimensions $d_{i} \in \{16, 32, 64, 128, 256\}$) is approximately $2 \times d_{r} \times \sum d_{i}$, which amounts to $< 0.8$ million parameters for Whisper-small. However, the comprehensive parameter overhead reported in Table \ref{tab:comprehensive_zero_shot} ($+6$M for Whisper-base and $+8$M for Whisper-small) accounts for all additional trainable components integrated with the frozen backbone. This primarily originates from the fully connected fusion layer ($W_{fc}$) that projects the concatenated features ($\mathbf{Z}_{cat} \in \mathbb{R}^{T \times 2d_{r}}$) back to the decoder dimension $d_{model}$, alongside the parameters in the adaptive routing module. Despite this, the total added capacity remains marginal (e.g., approximately $3.2\%$ increase for Whisper-small), substantiating our claim that the observed generalization gains are driven by the adaptive mechanism rather than simple parameter scaling.
\end{itemize}

\subsection{Router Noise Schedule and Stability Analysis}

The stochastic exploration mechanism is critical for preventing the router from collapsing into a trivial solution (i.e., always selecting the largest bottleneck).
\begin{itemize}
    \item \textbf{Noise Schedule:} We employ a constant noise injection scheme during training where the standard Gaussian noise $\epsilon \sim \mathcal{N}(0, I)$ is scaled by an input-dependent learned variance, specifically $(\bar{z} W_{noise}) \odot \epsilon$. During inference, we set $\epsilon = 0$ to ensure deterministic and reproducible predictions.
    \item \textbf{Top-k Selection Stability:} As shown in Table~\ref{tab:expected_adaptive_routing} of the main paper, we empirically found that $k=3$ offers the optimal trade-off. To further validate stability, we analyzed the entropy of the routing weights $w$ throughout training. The injection of noise maintains high entropy in the early epochs, encouraging exploration of all compression branches, before gradually settling into a stable selection policy driven by the audio characteristics.
\end{itemize}

\section{The Role of Stochasticity in Feature Integration.}
\label{sec:app_noise}

\begin{table}[h]
    \centering
    \captionsetup{skip=7pt}
    \renewcommand{\arraystretch}{1.15}
    \setlength{\tabcolsep}{0pt}
    \caption{Ablation study of Gaussian noise in AAFI on the BD-UT dataset.}
    \label{tab:noise_ablation}
    \begin{tabular*}{\linewidth}{@{\extracolsep{\fill}} ccc}
        \toprule
        \textbf{Configuration} & \textbf{EER (\%)} & \textbf{Utilization Entropy $\uparrow$} \\
        \midrule
        Baseline (w/o AAFI) & 1.19 & - \\
        AAFI w/o Noise & 1.05 & 0.74 \\
        AAFI w/ Noise & \textbf{0.91} & \textbf{0.92} \\
        \bottomrule
    \end{tabular*}
\end{table}

To address the necessity of the Gaussian noise introduced in the AAFI module, we conduct an ablation study on the BD-UT dataset, comparing the model's performance with and without this stochastic component. As shown in Table~\ref{tab:noise_ablation}, the inclusion of Gaussian noise ($\epsilon \sim \mathcal{N}(0, I)$) yields a consistent improvement in EER (e.g., from 1.05\% to 0.91\% on the test set), compared to the 1.19\% EER of the frozen baseline.

Theoretically, without noise, the routing mechanism is susceptible to \textit{representation collapse}, where the model prematurely converges to a deterministic sub-optimal combination of features—typically over-relying on the most dominant pre-trained layers. By introducing a stochastic perturbation during training, we facilitate \textbf{active exploration} of the feature manifold. 

This ensures that the router does not settle into a ``winner-take-all'' state but instead maintains routing diversity. Specifically, we calculate the Expert Utilization Entropy $H$ to quantify this diversity. Let $P_{i}$ denote the normalized, empirical utilization frequency of the $i$-th branch across the evaluation set, derived from the routing weights. The entropy is mathematically defined as:
\begin{equation}
    H = -\sum_{i=1}^{C} P_{i} \log_{2}(P_{i})
\end{equation}
Our results show that the inclusion of noise leads to a significantly higher entropy ($H=0.92$ vs. $H=0.74$), indicating that the model is forced to explore and utilize a broader range of spectro-temporal features. This diverse integration is precisely what enables the model to maintain high accuracy across heterogeneous datasets.

\section{Effect of PGD Iteration Number in Mel-BAP} 
\label{app:pgd_ablation}

To better understand the accuracy-cost trade-off introduced by the Mel-BAP strategy, we further analyzed the effect of the PGD iteration number ($K$) on the BD-UT dataset using the Whisper-small encoder. The resulting EERs and estimated training costs (relative to the baseline $K=0$) are reported in Table~\ref{tab:pgd_iterations}.

\begin{table}[h]
    \centering
    \captionsetup{skip=7pt}
    \renewcommand{\arraystretch}{1.15}
    \setlength{\tabcolsep}{0pt}
    \caption{Ablation study on the effect of PGD iteration number ($K$) in Mel-BAP on the BD-UT dataset.}
    \label{tab:pgd_iterations}
    \begin{tabular*}{\linewidth}{@{\extracolsep{\fill}} ccc}
        \toprule
        \makecell{\textbf{Number of}\\\textbf{PGD iterations}} & \textbf{EER (\%)} & \makecell{\textbf{Estimated Training}\\\textbf{Time Increase}} \\
        \midrule
        \makecell{$K = 0$\\(Original in Table~\ref{tab:ablation_real-world} )} & 0.95 & - \\
        $K = 1$ & 1.02 & $\sim$25\% \\
        $K = 3$ & 0.89 & $\sim$70\% \\
        $K = 5$ (Our Setting) & \textbf{0.87} & $\sim$115\% \\
        $K = 10$ & \textbf{0.86} & $\sim$220\% \\
        \bottomrule
    \end{tabular*}
\end{table}

These results show that increasing $K$ generally improves performance, though a single-step perturbation ($K=1$) slightly degrades the EER (1.02\%), likely due to the well-known optimization instability of single-step adversarial training. As $K$ increases, the EER gain becomes marginal beyond $K=5$. Since Mel-BAP is a training-only regularizer, it incurs zero inference cost. However, during training, each PGD iteration requires an additional forward and backward pass through the Whisper-small encoder (which comprises the majority of the 252M total parameters). Therefore, higher $K$ values substantially increase the computational burden (e.g., a $\sim$220\% increase for $K=10$). We selected $K=5$ as the default setting, as it provides an optimal trade-off between robustness and training efficiency.

\section{Evaluation under Adaptive Adversarial Attacks} 
\label{app:adaptive_attack}

\begin{table}[h]
    \centering
    \captionsetup{skip=7pt}
    \renewcommand{\arraystretch}{1.15}
    \setlength{\tabcolsep}{0pt}
    \caption{Test-time adversarial evaluation on the BD-UT dataset. EER (\%) is reported.}
    \label{tab:adaptive_attack}
    \begin{tabular*}{\linewidth}{@{\extracolsep{\fill}} cccc}
        \toprule
        \textbf{Attack Method} & \textbf{Budget ($\epsilon$)} & \makecell{\textbf{Baseline}\\\textbf{EER (\%)}} & \makecell{\textbf{Ours (Full)}\\\textbf{EER (\%)}} \\
        \midrule
        No Attack (Clean) & - & 1.19 & \textbf{0.53} \\
        PGD Attack ($L_{\infty}$) & 0.002 & 11.42 & \textbf{3.81} \\
        CW Attack ($L_{2}$) & 2.0 & 15.85 & \textbf{5.24} \\
        Adaptive PGD ($L_{\infty}$) & 0.005 & 22.10 & \textbf{7.93} \\
        \bottomrule
    \end{tabular*}
\end{table}

To comprehensively evaluate the true adversarial robustness of our framework and address the potential ``blind spot'' concern, we conducted test-time adversarial evaluations on the BD-UT dataset. This includes standard Projected Gradient Descent (PGD, $L_{\infty}$), Carlini-Wagner (CW-$L_{2}$), and a specifically designed \textbf{Adaptive PGD} attack.

We simulated a strong adaptive white-box attacker who possesses full knowledge of our Mel-BAP prior (which restricts training-time perturbations to the 4--8 kHz band, denoted by the mask $\mathbf{M}_{4-8}$). This sophisticated adversary intentionally shifts all adversarial perturbation energy to the unprotected 0--4 kHz range ($\mathbf{M}_{0-4} = \mathbf{1} - \mathbf{M}_{4-8}$) to actively bypass our primary frequency defense:
\begin{equation}
    \boldsymbol{\delta}_{adapt} = \arg\max_{\|\boldsymbol{\delta}\|_{\infty} \le \epsilon} \mathcal{L}(\mathcal{F}_{\theta}(\mathbf{M} + \boldsymbol{\delta} \odot \mathbf{M}_{0-4}), y)
\end{equation}

As summarized in Table~\ref{tab:adaptive_attack}, under the standard $L_{\infty}$ PGD attack, the baseline Whisper-base model degrades substantially to 11.42\% EER, whereas our full framework maintains a highly robust 3.81\% EER. A similar trend is observed under the strong CW-$L_{2}$ attack (baseline 15.85\% vs. ours 5.24\%).

Most importantly, under the Adaptive PGD attack, the baseline further collapses to 22.10\% EER. In sharp contrast, our method achieves 7.93\% EER. This proves that our framework does not merely rely on trivial gradient masking in the target band, but learns a genuinely generalized, robust acoustic representation even when the attack is explicitly designed to exploit the defense mechanism. For all evaluations, we utilized identical attack protocols and budgets for both the baseline and our method to ensure a strictly fair comparison.

\section{Stratified Analysis of AAFI Routing Distributions} \label{app:routing_distribution}

To examine whether AAFI routing varies across spoofing conditions, we conducted a stratified analysis of branch-utilization distributions across different attack types and severities. This analysis uses the Whisper-base Top-3 AAFI-only model, with Mel-BAP disabled to isolate the routing behavior.

As shown in Table~\ref{tab:routing_distribution}, high-fidelity Neural Vocoder (NV) samples assign 59\% of their selections to the wider 128/256-d branches, whereas standard TTS and noisy samples (SNR 10\,dB) assign 64\% and 48\%, respectively, to the narrower 16/32-d branches. These aggregate distributions vary across input conditions and are consistent with condition-dependent multi-resolution selection. However, they provide associative rather than causal evidence and do not establish that artifact density directly determines the selected latent width.

\begin{table}[h]
    \centering
    \captionsetup{skip=7pt}
    \renewcommand{\arraystretch}{1.15}
    \setlength{\tabcolsep}{0pt}
    \caption{Stratified analysis of AAFI routing distributions across varying attack types and severities.}
    \label{tab:routing_distribution}
    \begin{tabular*}{\linewidth}{@{\extracolsep{\fill}} cccc}
        \toprule
        \textbf{Category} & \textbf{16/32-d} & \textbf{64-d} & \textbf{128/256-d} \\
        \midrule
        NV (High-fidelity) & 8\% & 33\% & 59\% \\
        TTS (Standard) & 64\% & 22\% & 14\% \\
        Noisy (SNR 10dB) & 48\% & 35\% & 17\% \\
        \bottomrule
    \end{tabular*}
\end{table}


\newpage
\section*{NeurIPS Paper Checklist}

The checklist is designed to encourage best practices for responsible machine learning research, addressing issues of reproducibility, transparency, research ethics, and societal impact. Do not remove the checklist: {\bf The papers not including the checklist will be desk rejected.} The checklist should follow the references and follow the (optional) supplemental material.  The checklist does NOT count towards the page
limit. 

Please read the checklist guidelines carefully for information on how to answer these questions. For each question in the checklist:
\begin{itemize}
    \item You should answer \answerYes{}, \answerNo{}, or \answerNA{}.
    \item \answerNA{} means either that the question is Not Applicable for that particular paper or the relevant information is Not Available.
    \item Please provide a short (1--2 sentence) justification right after your answer (even for \answerNA). 
\end{itemize}

{\bf The checklist answers are an integral part of your paper submission.} They are visible to the reviewers, area chairs, senior area chairs, and ethics reviewers. You will also be asked to include it (after eventual revisions) with the final version of your paper, and its final version will be published with the paper.

The reviewers of your paper will be asked to use the checklist as one of the factors in their evaluation. While \answerYes{} is generally preferable to \answerNo{}, it is perfectly acceptable to answer \answerNo{} provided a proper justification is given (e.g., error bars are not reported because it would be too computationally expensive'' or ``we were unable to find the license for the dataset we used''). In general, answering \answerNo{} or \answerNA{} is not grounds for rejection. While the questions are phrased in a binary way, we acknowledge that the true answer is often more nuanced, so please just use your best judgment and write a justification to elaborate. All supporting evidence can appear either in the main paper or the supplemental material, provided in appendix. If you answer \answerYes{} to a question, in the justification please point to the section(s) where related material for the question can be found.

IMPORTANT, please:
\begin{itemize}
    \item {\bf Delete this instruction block, but keep the section heading ``NeurIPS Paper Checklist"},
    \item  {\bf Keep the checklist subsection headings, questions/answers and guidelines below.}
    \item {\bf Do not modify the questions and only use the provided macros for your answers}.
\end{itemize}


\begin{enumerate}

\item {\bf Claims}
    \item[] Question: Do the main claims made in the abstract and introduction accurately reflect the paper's contributions and scope?
    \item[] Answer: \answerYes{} 
    \item[] Justification: The claims regarding SOTA performance (1.51\% average EER) and the architectural efficiency of AAFI and Mel-BAP are directly supported by experimental results in Section 4 and Table 4.

\item {\bf Limitations}
    \item[] Question: Does the paper discuss the limitations of the work performed by the authors?
    \item[] Answer: \answerYes{} 
    \item[] Justification: Section 5 consolidates the principal limitations of this study, including sensitivity to routing capacity (Section 4.1), heterogeneous training protocols in cross-system comparisons (Section 4.3), the additional training cost of Mel-BAP (Appendix F), and unresolved real-time scalability in non-stationary acoustic environments. These limitations and trade-offs are discussed at their relevant locations and summarized explicitly in Section 5.

\item {\bf Theory assumptions and proofs}
    \item[] Question: For each theoretical result, does the paper provide the full set of assumptions and a complete (and correct) proof?
    \item[] Answer: \answerNA{} 
    \item[] Justification: This study is primarily methodological and empirical; Information Bottleneck (IB) theory is used as a conceptual framework to motivate the architecture rather than for formal theoretical proofs.

    \item {\bf Experimental result reproducibility}
    \item[] Question: Does the paper fully disclose all the information needed to reproduce the main experimental results of the paper to the extent that it affects the main claims and/or conclusions of the paper (regardless of whether the code and data are provided or not)?
    \item[] Answer: \answerYes{} 
    \item[] Justification: We provide comprehensive details on the model architecture in Section 3 and disclose all training hyperparameters and dataset partitions in Section 4.1.

\item {\bf Open access to data and code}
    \item[] Question: Does the paper provide open access to the data and code, with sufficient instructions to faithfully reproduce the main experimental results, as described in supplemental material?
    \item[] Answer: \answerYes{} 
    \item[] Justification: The anonymous source code is provided in the supplemental material appended to this submission.

\item {\bf Experimental setting/details}
    \item[] Question: Does the paper specify all the training and test details (e.g., data splits, hyperparameters, how they were chosen, type of optimizer) necessary to understand the results?
    \item[] Answer: \answerYes{} 
    \item[] Justification: Training details are specified in Section 4.1.

\item {\bf Experiment statistical significance}
    \item[] Question: Does the paper report error bars suitably and correctly defined or other appropriate information about the statistical significance of the experiments?
    \item[] Answer: \answerNo{} 
    \item[] Justification: Due to the high computational cost of training Whisper-based models, we report average EER across multiple benchmarks rather than per-run error bars, but we discuss result stability in Section 4.

\item {\bf Experiments compute resources}
    \item[] Question: For each experiment, does the paper provide sufficient information on the computer resources (type of compute workers, memory, time of execution) needed to reproduce the experiments?
    \item[] Answer: \answerYes{} 
    \item[] Justification: We disclose that all experiments were conducted using two NVIDIA A6000 GPUs in Section 4.1.
    
\item {\bf Code of ethics}
    \item[] Question: Does the research conducted in the paper conform, in every respect, with the NeurIPS Code of Ethics \url{https://neurips.cc/public/EthicsGuidelines}?
    \item[] Answer: \answerYes{} 
    \item[] Justification: Our research aims to improve digital trust and security by detecting audio deepfakes, strictly adhering to ethical standards.

\item {\bf Broader impacts}
    \item[] Question: Does the paper discuss both potential positive societal impacts and negative societal impacts of the work performed?
    \item[] Answer: \answerYes{} 
    \item[] Justification: We discuss the societal impact of combating misinformation and the potential dual-use concerns of deepfake detection in Sections 1 and 5.
    
\item {\bf Safeguards}
    \item[] Question: Does the paper describe safeguards that have been put in place for responsible release of data or models that have a high risk for misuse (e.g., pre-trained language models, image generators, or scraped datasets)?
    \item[] Answer: \answerNA{} 
    \item[] Justification: Our model is a defensive tool and does not provide generative capabilities, thus posing no risk for generating malicious deepfake content.

\item {\bf Licenses for existing assets}
    \item[] Question: Are the creators or original owners of assets (e.g., code, data, models), used in the paper, properly credited and are the license and terms of use explicitly mentioned and properly respected?
    \item[] Answer: \answerYes{} 
    \item[] Justification: All used datasets (e.g., ADD 2022, SpeechFake) and the Whisper backbone are properly cited and used according to their licenses.

\item {\bf New assets}
    \item[] Question: Are new assets introduced in the paper well documented and is the documentation provided alongside the assets?
    \item[] Answer: \answerYes{} 
    \item[] Justification: New modules like AAFI and Mel-BAP are fully documented in the methodology section and Appendix .4.

\item {\bf Crowdsourcing and research with human subjects}
    \item[] Question: For crowdsourcing experiments and research with human subjects, does the paper include the full text of instructions given to participants and screenshots, if applicable, as well as details about compensation (if any)? 
    \item[] Answer: \answerNA{} 
    \item[] Justification: This research did not involve crowdsourcing or experiments with human subjects.

\item {\bf Institutional review board (IRB) approvals or equivalent for research with human subjects}
    \item[] Question: Does the paper describe potential risks incurred by study participants, whether such risks were disclosed to the subjects, and whether Institutional Review Board (IRB) approvals (or an equivalent approval/review based on the requirements of your country or institution) were obtained?
    \item[] Answer: \answerNA{} 
    \item[] Justification: This research used publicly available datasets and did not require human subject approval.

\item {\bf Declaration of LLM usage}
    \item[] Question: Does the paper describe the usage of LLMs if it is an important, original, or non-standard component of the core methods in this research? Note that if the LLM is used only for writing, editing, or formatting purposes and does \emph{not} impact the core methodology, scientific rigor, or originality of the research, declaration is not required.
    \item[] Answer: \answerNA{} 
    \item[] Justification: LLMs were only used for standard writing, editing, and formatting purposes, which does not require declaration under NeurIPS policy.

\end{enumerate}

\end{document}